\documentclass{PoS}

\title{Critical structure of the QCD medium}

\ShortTitle{Critical structure of the QCD medium}

\author{\speaker{Bernd-Jochen
    Schaefer}\\
  Institut f\"ur Physik, Karl-Franzens-Universit\"at, A-8010
  Graz, Austria\\
  E-mail: \email{bernd-jochen.schaefer@uni-graz.at}}

\abstract{Fluctuations in the vicinity of a phase transition are
  important but neglected in mean-field theory. In order to assess the
  influence of such fluctuations on the critical endpoint and the size
  of the critical region in the QCD phase diagram, a mean-field
  calculation of a two-flavor quark-meson model is compared with a
  renormalization group approach. However, due to the lack of
  confinement in this effective model the equation of state near the
  chiral phase transition is still unrealistic. A first improvement of
  this model can be achieved by coupling quark degrees of freedom to
  the Polyakov loop, consequently incorporating certain aspects of
  confinement. The influence of these modifications on the resulting
  phase diagram is discussed.}

\FullConference{Critical Point and Onset of Deconfinement
          4th International Workshop\\
		 July 9-13 2007\\
		 GSI Darmstadt,Germany}

\begin{document}

\section{Introduction}

Driven by the future heavy-ion programs of CBM at the GSI FAIR
facility, and soon also at the CERN LHC, the properties of strongly
interacting matter at finite temperature $T$ and baryon number density
are of growing interest. On the theoretical side QCD at high
temperature and/or quark chemical potential $\mu$ predicts a phase
transition from the ordinary hadronic phase to a chirally restored and
deconfined quark gluon plasma (QGP) phase. Based on calculations in
effective models, as well as universality arguments, the order of the
phase transition depends on the number of quark flavors and on the
value of the quark masses. For physical values of the quark masses and
small baryon densities the transition as function of temperature is
continuous. On the other hand, for small temperature the transition as
function of the chemical potential is of first-order. This suggests
the existence of a critical endpoint (CEP) in the phase diagram, given
by the endpoint of a first-order transition line in the
$(T,\mu)$-plane \cite{Asakawa1989, Barducci1989, Berges1999a}. The
phase transition at the CEP is of second-order and belongs to the
three-dimensional Ising universality class. At this point the QCD free
energy has a genuine singularity and as a consequence both the chiral
and the quark number susceptibility diverge. Due to, e.g., the large
strange quark-mass sensitivity the precise location of this endpoint
is not yet known.

From QCD lattice simulations for vanishing $\mu$, it is known that the
transition temperature in the presence of light quarks is lowered
substantially from its value in the pure gauge limit of infinitely
heavy quarks. Furthermore, in the limit of vanishing up- and down
quark masses as well as infinite strange quark mass, the chiral
transition is likely to be of second-order and its static critical
behavior falls in the universality class of the Heisenberg $O(4)$
model in three dimensions \cite{Pisarski1984}. But the situation on
the lattice for finite $\mu$ is much less clear due to the notorious
fermion sign problem. Nevertheless, from direct numerical evaluations
of the QCD partition function or also from a Taylor expansion of the
pressure around vanishing $\mu$ some evidence exists for a CEP in the
phase diagram at finite $\mu$ \cite{Fodor2002, Forcrand2002,
  Allton2003}. For two-flavor massless QCD the CEP turns into a
tricritical point (TCP) where for small $\mu$'s the $O(4)$ line of
critical points terminates. For larger $\mu$'s the transition is again
of first-order.

In order to interpret the physical content of these lattice findings
they have to be compared to model studies. A variety of model studies
of the CEP and the critical region are available in the literature,
but most of them are based on a mean-field description of the phase
transition, consequently neglecting quantum fluctuations. However, it
is well-known, especially in the vicinity of a phase transition, that
fluctuations become more and more important and mean-field theory
fails to describe adequately the critical behavior of phase
transitions. Therefore, it is necessary to go beyond mean-field theory
to arrive at a proper description. An efficient way, to go beyond
mean-field theory, is the renormalization group (RG) method which
considers the universal and non-universal aspects not only of
second-order but also of first-order phase transitions \cite{RG,
  RG_qm}.

In this contribution we summarize recent results on the size of the
critical region around the TCP and CEP obtained with an effective
two-flavor quark-meson model, both for a mean-field approximation and
a RG approach \cite{Schaefer2007}. Due to the lack of confinement in
this model single quark states are already excited in the chirally
broken phase yielding an unrealistic equation of state (EoS) near the
phase transition \cite{Jungnickel1996, Schaefer1999}. Coupling quark
degrees of freedom to the Polyakov loop certain aspects of confinement
are incorporated that improve the EoS \cite{Fukushima2003, Ratti2006,
  Megias2006, Sasaki2007}. Finally, the modifications caused by the
Polyakov loop on the phase diagram is presented and compared with the
one obtained in the pure quark-meson model \cite{Schaefer2007a}.

\section{The size of the critical region around the TCP and CEP}

The size of a critical region around a critical endpoint is defined
through the breakdown of mean-field theory and emergence of nontrivial
critical exponents. The size usually can be determined by the
well-known Ginzburg criterion which is based on an expansion of the
singular part of the free energy for a second-order phase transition.
However, since the expansion coefficients are not known for the strong
interaction the Ginzburg criterion is only of limited use in the
present context. Universality arguments are also not helpful if the
underlying microscopic dynamics is not well determined. For example,
in the cases of the $\lambda$-transition of liquid He$^4$ and the
superconducting transition of metals, both transitions lie in the same
universality class but their critical regions defined by their
corresponding Ginzburg-Levanyuk temperature deviate from each other by
several orders of magnitude.

An estimate for the size of the critical region for hadronic matter
also can be defined by calculating the in-medium scalar and chiral
static susceptibility using their enhancement as the criterion
\cite{Hatta2003}. In general, static susceptibilities are obtained
from the dynamic response function $\chi_{ab} (\omega, \vec q)$ in the
static ($\omega=0$) and long wavelength limit ($\vec q \to 0$) where
$a, b$ denote some external fields. The scalar static susceptibility
$\chi_\sigma$ corresponds to the zero-momentum projection of the
scalar propagator which encodes all fluctuations of the chiral order
parameter $\langle \bar q q \rangle$. Thus, the maximum of
$\chi_\sigma$ as function of temperature or quark chemical potential
should coincide with the most rapid change in the chiral order
parameter. It is related to the sigma meson mass via
$\chi_\sigma \sim M_\sigma^{-2}$. Similarly, the chiral or quark
number susceptibility $\chi_q$ is the response of the net quark number
density $n_q$ to an infinitesimal variation of the quark chemical
potential, $\chi_q = \partial n_q/\partial \mu_q$.

In mean-field approximation and for a physical pion mass the
two-flavor quark-meson model exhibits a smooth crossover on the
temperature axis and a first-order chiral phase transition on the
density axis \cite{Scavenius2001}. For increasing temperatures the
first-order transition line terminates at the CEP. Along the line of a
first-order phase transition the thermodynamic potential has two
minima of equal depth which are separated by finite potential barrier.
The height of the barrier is largest at zero temperature and finite
chemical potential and decreases towards higher temperature. At the
CEP the barrier and accordingly the latent heat of the transition
disappears and the potential flattens. At this point the phase
transition is of second-order and characterized by long-wavelength
fluctuations of the order parameter which is in our case proportional
to the scalar $\sigma$-field. As a consequence the scalar sigma mass
must vanish at this point which can be seen in the behavior of the
in-medium meson masses: in the vicinity of the CEP the sigma mass as
function of temperature and quark chemical potential drops below the
pion mass which stays always finite since the chiral symmetry is still
explicitly broken. For temperatures and chemical potentials above the
chiral transition the sigma mass increases again and will degenerate
with the pion mass signaling restoration of chiral symmetry.

At the CEP the slope of the quark number density tends to infinity
which will yield a diverging susceptibility exactly at this point. For
temperatures below the critical one the quark number density jumps
because of the first-order phase transition. For temperatures above
the CEP the discontinuity vanishes across the transition line and the
density changes gradually due to the smooth crossover. This finally
produces a finite height of the quark number susceptibility $\chi_q$.
Thus, in equilibrium $\chi_q$ diverges only at the CEP and is finite
everywhere else\footnote{This changes for non-equilibrium systems:
  when entering the coexisting region of the first-order line, the
  susceptibility also diverges along the isothermal spinodal line,
  see~\cite{Sasaki:2007db} for further details.}. The height of
$\chi_q$ decreases for decreasing chemical potentials towards the
$T$-axis. For temperatures below the CEP $\chi_q$ is discontinuous and
jumps across the first-order line. In the vicinity of the CEP the
quark number density is always finite but the susceptibility becomes
large. Since the quark number susceptibility is proportional to the
isothermal compressibility $\kappa_T$ via the relation
$\kappa_T = \chi_q/n_q^2$ this behavior indicates that the system is
easy to compress around the critical point.

\begin{figure}
  \centering  
  \includegraphics[width=0.45\textwidth,angle=-90]{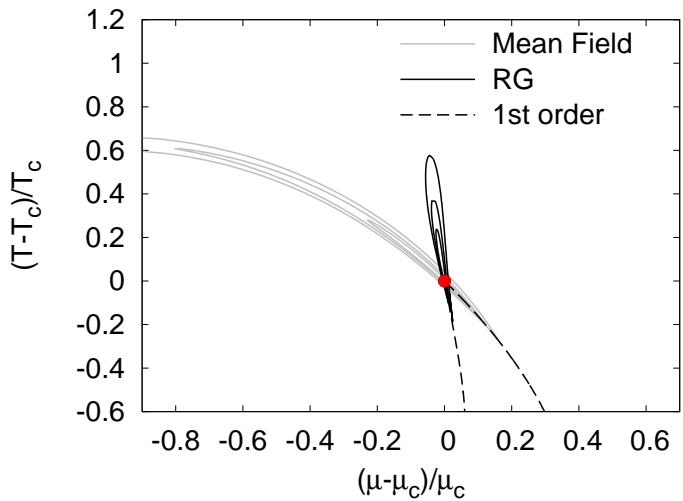} 
  \caption{The contour regions in the phase diagram for three
    different ratios of the scalar susceptibilities
    $R_s = \chi_\sigma (T,\mu)/\chi_\sigma (0,0)$ ($R_s = 10,15,20$)
    around the CEP in reduced units.}
\label{fig:cr_CEP} 
\end{figure}

Figure \ref{fig:cr_CEP} shows a contour plot of the scalar
susceptibility divided by the vacuum susceptibility
\[ 
R_s (T,\mu) = \frac{ \chi_\sigma (T,\mu)}{\chi_\sigma (0,0)}
\]
for three fixed ratios around the CEP in the phase diagram in reduced
units. The light curves are the mean-field and the other ones the RG
results which we will discuss later. The region of the enhanced
susceptibility is elongated in the direction of the extrapolated
first-order transition line. The deeper reason for this shape can be
understood by a study of the critical exponents of the susceptibility
which specify its power-law singularity. In the case of the
susceptibility the form of this divergence depends on the path by
which one approaches the critical point. For the path asymptotically
parallel to the first-order transition line the divergence scales with
an exponent $\gamma$ which in mean field is $\gamma=1$. For any other
path, not parallel to the first-order line, the divergence scales with
another exponent $\epsilon$ which in mean-field theory is equal to
$2/3$. Since $\gamma > \epsilon$ the susceptibility is enhanced in the
direction parallel to the first-order transition line. This is the
reason for the elongated shape of the critical region in the phase
diagram.

Universality arguments as well as lattice QCD simulations for two
quark flavors without $U_A(1)$ anomaly in the chiral limit predict at
vanishing quark chemical potential that the effective theory for the
chiral order parameter is the same as for the $O(4)$ model, which has
a second-order phase transition. It is expected that the static
critical behavior falls into the universality class of the
$O(4)$-symmetric Heisenberg model in three dimensions. When reducing
the pion mass by varying the explicit chiral symmetry breaking
parameter of our quark-meson model, the CEP moves towards the
$T$-axis. Already for the pion mass $M_\pi \sim 70$ MeV the CEP
disappears and chiral symmetry is restored via a first-order
transition for all temperatures and quark chemical potentials. As a
consequence this model does not have a tricritical point in the chiral
limit in contradiction to universality arguments and lattice
simulations \cite{Fujii2004}. But as already stated in \cite{Scavenius2001}, within the
mean-field approximation the order of the phase transition in the
chiral limit of the quark-meson model strongly depends on the values
for the model parameters. The way how to extrapolate towards the
chiral limit is not unique. Thus, the mean-field approximation fails
to properly describe the expected critical behavior in the chiral
limit at least for the parameter set chosen. 

This is remedied in the RG approach and a second-order phase
transition, which lies in the expected $O(4)$ universality class, is
found in the chiral limit at finite temperature \cite{Berges2003,
  Tetradis2003}. For finite chemical potential the second-order
transition ends in a TCP. For finite quark or pion masses this
transition is washed out and becomes a smooth crossover with a
critical endpoint. Thus, in the RG framework the relationship and the
correlations between the TCP and the various CEP's, obtained by
varying the pion mass, can be studied. In addition, the influence of
fluctuations on the susceptibilities and the critical region around
the CEP can be assessed.

\begin{figure}
  \centering
  \includegraphics[width=0.45\textwidth,angle=-90]{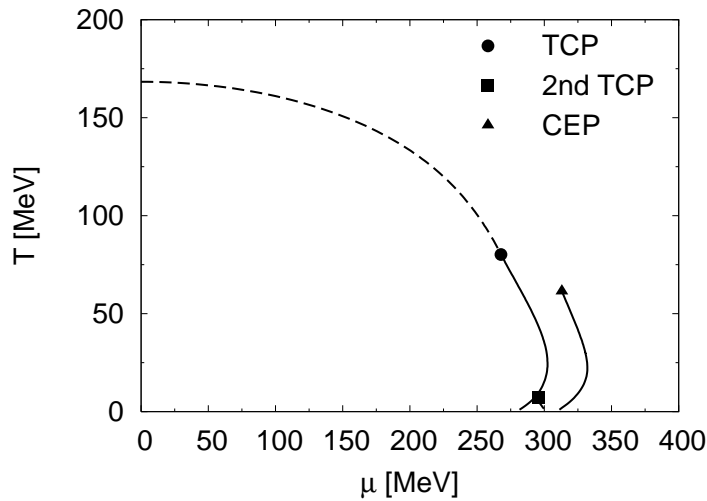}
  \caption{Two phase diagrams for the quark-meson model obtained with
    the RG: One for physical pion masses (right solid line which ends
    in the CEP) and another one for the chiral limit. Solid lines
    denote first-order and dashed lines second-order transition
    lines.}
  \label{fig:RG_phasediagram} 
\end{figure}

The resulting phase diagrams for the chiral limit and for physical
pion masses $M_\pi \sim 130$ MeV are both shown in
Fig.~\ref{fig:RG_phasediagram}. The location of the TCP for our choice
of parameters is at $T_c^t \sim 80$ MeV and $\mu_c^t \sim 270$ MeV.
For temperatures below the TCP the phase transition changes initially
to a first-order transition. For temperatures below $10$ MeV two phase
transitions with a second tricritical point emerge
\cite{Schaefer2005}. A larger constituent quark mass pushes the
location of the first TCP towards the temperature axis and the
location of the splitting point of the two phase transition lines down
towards the $\mu$-axis. All qualitative features of the two transition
lines survive, only the area bounded by the two transition lines is
reduced for increasing quark masses. Thus, in contrast to the
mean-field approximation the RG method yields a TCP in the phase
diagram for the chiral limit as expected.

As the pions become massive the TCP turns into a CEP, which lies in
the universality class of the three-dimensional Ising model. The
location of the CEP for physical pion masses is at $T_c^c \sim 62$ MeV
and $m_c^c \sim 313$ MeV. Compared to recent lattice and other model
studies the location of the TCP and consequently of the CEP is at
lower temperatures due to the omission of other degrees of freedom in
the used quark-meson model. Typical for the RG treatment is the
bending of the first-order transition lines for smaller temperatures.
But very close to the $\mu$-axis the slope of the first-order boundary
$dT/d\mu$ tends to infinity similar to the mean-field phase diagram
and is also in agreement with the Clausius-Clapeyron relation
\cite{Aigner2007}. In the chiral limit below the splitting point the
right second-order transition line turns into an crossover for finite
quark masses which is not visible in Fig.~\ref{fig:RG_phasediagram}.
Analogously, the second TCP should turn into a critical point whose
remnants can be seen in the order parameter and meson masses
\cite{Schaefer2005}.

The results of a recalculation of the contour plot around the CEP in
the framework of the RG approach is also shown in
Fig.~\ref{fig:cr_CEP}. The critical region is again elongated in the
direction of the first-order transition line, but it is now much more
compressed. While the interval of the critical region in the
temperature direction is comparable with the one obtained in the
mean-field approximation, the effect in the chemical potential
direction is dramatic. In the RG calculation the interval is shrunken
by almost one order of magnitude, despite the fact that the
corresponding critical exponents are quite similar. For example, at
the CEP the susceptibility diverges with the critical exponent
$\epsilon \sim 0.74$ which is consistent with the one of the expected
3D Ising universality class $\epsilon=0.78$. Thus, as a consequence of
fluctuations, the size of the critical region around the CEP is
substantially reduced as compared to the mean-field calculation. This
may also have consequences for the experimental localization of the
CEP in the phase diagram since it further complicates its detection
through event-by-event fluctuations.

\section{The quark-meson model with Polyakov-loop dynamics}

Despite the success of the RG approach in predicting the expected
critical behavior of the thermodynamics in the quark-meson model,
explicit gluonic degrees of freedom, which are known to play an
important role in the thermodynamics of QCD and are associated with
confinement aspects are missing in this model. One possibility to
incorporate such effects is the coupling of the quark-meson model to
the Polyakov loop. This results in an coupled effective
Polyakov--quark-meson (PQM) model with an interaction potential
between quarks, mesons and the Polyakov loop variables
$\phi$,$\bar \phi$. The PQM model includes the chiral aspects of QCD
as well as certain aspects of confinement.

The order parameter $\langle \phi \rangle$, and respectively
$\langle \bar \phi \rangle$, vanishes in the confined phase where the
free energy of a single heavy quark, respectively antiquark, diverges
and is finite in the deconfined phase. In the presence of dynamical
quarks, the free energy of a quark-antiquark pair does not diverge
anymore, and the order parameter is always non-vanishing. For finite
quark chemical potential the free energies of quarks and antiquarks
are different. Since $\langle \phi \rangle$ is related to the free
energy of quarks and the hermitian conjugate
$\langle \bar \phi \rangle$ to that of antiquarks, their modulus
differs in general.

In pure Yang-Mills theory the mean value $\langle \phi \rangle$,
$\langle \bar \phi \rangle$ are given by the minima of the effective
Polyakov-loop potential ${ U} (\phi ,\bar \phi )$. It can be
constructed from lattice data for the expectation values
\cite{Pisarski2000} or from a RG calculation \cite{BGP}. Here, a
polynomial expansion in $\langle \phi \rangle$,
$\langle \bar \phi \rangle$ up to quartic terms is used. The expansion
coefficients are fixed to reproduce thermodynamic lattice results for
the pure Yang-Mills sector. This potential has a first-order phase
transition at the critical temperature $T_0 = 270$ MeV. In the
presence of dynamical quarks, the running gauge coupling is changed
due to fermionic contributions. In our approximation to the
Polyakov-loop potential this only leads to a modification of the first
expansion coefficient $b_2$ in front of the quadratic fields. The size
of this effect can be estimated within perturbation theory. At zero
temperature it leads to an $N_f$-dependent decrease of
$\Lambda_{\rm QCD}$, which translates into an $N_f$-dependent decrease
of the critical temperature $T_0$ at finite temperature.
Table~\ref{tab:critt} shows the results for the $N_f$-dependent
critical temperature $T_0$ in the Polyakov-loop potential for massless
flavors. Massive flavors lead to a suppression factor in the
$\beta$-function of QCD which modifies $T_0$ further. E.g.~for $2+1$
flavors with a current strange quark mass $m_s \sim 150$ MeV a
$T_0(2+1) \sim 187$ MeV is obtained.

\begin{table}[h!]
  \begin{center} 
    \begin{tabular}{c||@{\hspace{2mm}}c@{\hspace{2mm}}|@{\hspace{2mm}}
        c@{\hspace{2mm}}|@{\hspace{2mm}}c@{\hspace{2mm}}|@{\hspace{2mm}}
        c@{\hspace{2mm}}|@{\hspace{2mm}}c@{\hspace{2mm}}}
      $N_f$ & $0$ & $1$ & $2$ & $2+1$ & $3$  \\
      \hline
      \hline
      $T_0$ [MeV] & 270 & 240 & 208 & 187 & 178 \\
      \hline
    \end{tabular}
    \caption{\label{tab:critt} The critical Polyakov-loop temperature $T_0$ for
      $N_f$ massless flavors.}
  \end{center}
\end{table}

A second step implements a $\mu$-dependent running coupling in the
$b_2$ coefficient, analogous to the $N_f$-dependence discussed above.
One can argue that this is a minimal necessary generalization because
without a $\mu$-dependent $b_2$ coefficient the
confinement-deconfinement phase transition has a higher critical
temperature than the chiral phase transition at vanishing chemical
potential. But this is an unphysical scenario because QCD with
dynamical massless quarks in the chirally restored phase cannot be
confining since the string breaking scale would be zero. 

As for the $N_f$-dependence one can resort to perturbative estimates,
by allowing for an additional $\mu$-dependent term in the one-loop
coefficient of the QCD $\beta$-function, which can be motivated by
using HTL/HDL results. This additional coefficient can be fixed such
that the chiral transition temperature and the
confinement-deconfinement transition agree at some arbitrary
non-vanishing $\mu$. Interestingly, it turns out that then the
transition temperatures agree for all values of $\mu$. This
$\mu$-dependence in the $\beta$-function then leads to a $T_0$ with an
additional $\mu$-dependence, such that $T_0 \to T_0 (\mu, N_f)$. Of
course, these novel modifications should be viewed as a rough estimate
of the $\mu$-dependence of $T_0$. For a more quantitative analysis the
non-perturbative running of the coupling in the presence of finite
temperature and quark density has to be considered. This can be
incorporated in a self-consistent RG-setting.

\begin{figure}
  \centering
  \includegraphics[width=0.45\textwidth,angle=-90]{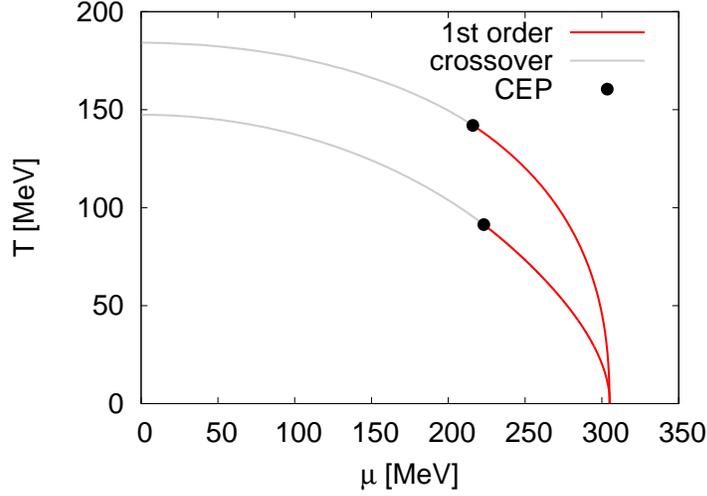}
  \caption{Phase diagram of the quark-meson model with Polyakov loop
    (upper lines) and without Polyakov loop dynamics (lower lines) in
    mean-field approximation.}
  \label{fig:cmp_phasediagram} 
\end{figure}

The phase structure of the PQM model is determined by the behavior of
the order parameters $\langle \sigma \rangle$, $\langle \phi\rangle$
and $\langle \bar \phi \rangle$ and the grand canonical potential as a
function of temperature and quark chemical potential. The phase
diagram in the $(T,\mu )$-plane resulting from the two flavor PQM
models in mean-field approximation is shown in
Fig.~\ref{fig:cmp_phasediagram} (upper lines). The bottom lines in
this figure display the phase diagram of the pure quark-meson model
without the Polyakov loop dynamic.

With the Polyakov loop modifications a chiral crossover temperature
$T_c \sim 184$ MeV at $\mu=0$ is found with an uncertainty of
$\sim 14$ MeV originating in the error estimate $\pm30$ MeV for $T_0$.
As an example for the estimate of the error, the two-loop running of
the gauge coupling leads to a $T_0 \sim 192$ MeV and hence a
$T_c \sim 177$ MeV. In the presence of dynamical quarks the Polyakov
loop shows also a crossover at the same pseudocritical temperature
which can be read off from the peak position of
$\partial \langle \bar q q \rangle /\partial T$ and
$\partial \langle \phi \rangle/ \partial T$.

Recently, a recalculation of the transition temperature on different
lattices for two light and one heavier quark mass close to their
physical values yields inconsistent results: On the one hand using the
Sommer parameter $r_0$ for the continuum extrapolation a
$T_c = 192\pm7$ MeV is found \cite{Cheng2006} and on the other hand in
another analysis with four different sets of lattice sizes
$N_\tau = 4,6,8$ and $10$ a $T_c = 151 \pm 3$ MeV is obtained
\cite{Aoki2006}. Within an Functional RG approach a critical value of
$T_c = 172$ MeV \cite{Braun2007} is achieved which again agrees with a
$T_c = 173\pm8$ MeV obtained in former two-flavor lattice simulations
with improved staggered fermions extrapolated to the chiral limit
\cite{Karsch2002}. Using the same parameters for the quark-meson model
without the Polyakov-loop modifications a crossover temperature of
$T_c \sim 150$ MeV emerges \cite{Schaefer2007, Schaefer2005}. This
temperature gap calls for refined studies both on the lattice as well
as analytical methods to resolve this discrepancy.

With and without the Polyakov loop modifications the phase diagram
features a critical endpoint (CEP), where the line of first-order
transitions terminates in a second-order transition. Lattice
simulations are not conclusive concerning the existence and location
of the CEP. There are indications from lattice simulations at small
chemical potentials that deconfinement and chiral symmetry restoration
appear along the same critical line in the phase diagram. For the PQM
model with an $\mu$-independent $T_0$ the coincidence of deconfinement
and chiral transition at $\mu=0$ disappears for finite $\mu$. The
deconfinement temperature is larger than the corresponding chiral
transition temperature. This is an unphysical scenario because the
deconfinement temperature should be smaller or equal to the chiral
transition temperature. Contrarily, with the $\mu$-dependent $T_0$
coinciding transition lines for the entire phase diagram within an
accuracy of $\pm5$ MeV are found.


\section{Summary}

The phase diagram of hadronic matter is analyzed in the two-flavor
quark-meson model by means of a Wilsonian RG approach. This model
captures essential features of QCD such as chiral symmetry breaking in
the vacuum and can therefore yield valuable insight into the critical
behavior associated with chiral symmetry. Of special importance is the
emergence of a CEP and the size of the critical region around the CEP
in connection with fluctuation signals in heavy-ion collisions. Most
studies of this issue have been performed in the mean-field
approximation which neglects thermal and quantum fluctuations. These
can be assessed in the RG approach which is able to correctly predict
critical exponents in the vicinity of critical points of the phase
diagram. 

In a mean-field calculation no TCP is found for the chosen parameter
set while the RG predicts its existence as is expected from
universality arguments. Because of the Gaussian fixed point structure
at the TCP mean-field exponents are expected what we also could
verify. When effects of finite current quark masses (or equivalently
finite pion masses) are included, a CEP emerges in both the mean-field
and RG calculation. By analyzing the scalar- and quark number
susceptibilities with the RG approach we found nontrivial critical
exponents which are consistent with the expected 3D Ising universality
class. As a consequence of fluctuations the size of the critical
region around the CEP is substantially reduced as compared to the
mean-field results. This is particularly true in the $\mu$-direction.

One of the truncations of the quark-meson model is the lack of
explicit gluonic degrees of freedom. This is addressed by the
introduction of the PQM model that includes certain aspects of gluon
dynamics via the Polyakov loop and represents a minimal synthesis of
the two basic principles of QCD at low temperatures: spontaneous
chiral symmetry breaking and confinement.

A limited set of input parameters is adjusted to reproduce lattice QCD
results in the pure gauge sector and pion properties in the hadron
sector. Then the PQM model correctly describes the step from the
first-order deconfinement transition observed in pure-gauge lattice
QCD with a $T_c \sim 270$ MeV to the crossover phenomenon with a
pseudocritical $T_c$ around 200 MeV when two light quark flavors are
added. The non-trivial result is that the crossovers for chiral
symmetry restoration and deconfinement almost coincide at small $\mu$
similar to lattice simulations. Via RG arguments it is possible to
estimate an $N_f$- and $\mu$-dependence in the parameters of the
Polyakov loop potential: the critical temperature of the Polyakov loop
model decreases with increasing $N_f$ and $\mu$. These modifications
yield coinciding peaks in the temperature derivative of the Polyakov
loop expectation value and the chiral condensate at $\mu=0$.
Interestingly, this coincidence of the deconfinement and chiral
symmetry restoration persists at finite $\mu$. These findings provide
a promising starting point for a functional RG study in the PQM model,
and further extensions towards full QCD.

\section*{Acknowledgment}

The work presented here was done in collaboration with Jochen Wambach.
The study on the PQM model was done also in collaboration with Jan
M.~Pawlowski. The author is grateful to the Technical University of
Darmstadt and GSI for the hospitality extended to him. This work is
supported by the BMBF Grant No. 06DA116.

\end{document}